\documentclass[review]{elsarticle}
\usepackage{lineno,hyperref}
\usepackage[english]{babel}
\usepackage[numbers]{natbib}
\usepackage{xcolor}
\usepackage{latexsym,amsmath,amssymb,amsbsy,graphicx,geometry}
\modulolinenumbers[5]

\begin{document}
	
	\begin{frontmatter}

\title{Kinetics of inherent processes counteracting crystallization in supercooled monatomic liquid}

\author[kfu,ufrc]{B.N. Galimzyanov\corref{cor1}}
\cortext[cor1]{Corresponding author}
\ead{bulatgnmail@gmail.com}

\author[kfu]{D.T. Yarullin}

\author[kfu,ufrc]{A.V. Mokshin}
\ead{anatolii.mokshin@mail.ru}

\address[kfu]{Kazan Federal University, 420008 Kazan, Russia} 
\address[ufrc]{Udmurt Federal Research Center of the Ural Branch of the Russian Academy of Sciences, 426067 Izhevsk, Russia}

\begin{abstract}
Crystallization of supercooled liquids is mainly determined by two competing processes associated with the transition of particles (atoms) from liquid phase to crystalline one and, vice versa, with the return of particles from crystalline phase to liquid one. The quantitative characteristics of these processes are the so-called attachment rate $g^{+}$ and the detachment rate $g^{-}$, which determine how particles change their belonging from one phase to another. In the present study, a {\it correspondence rule} between the rates $g^{+}$ and $g^{-}$ as functions of the size $N$ of growing crystalline nuclei is defined for the first time. In contrast to the well-known detailed balance condition, which relates $g^{+}(N)$ and $g^{-}(N)$ at $N=n_c$ (where $n_c$ is the critical nucleus size) and is satisfied only at the beginning of the nucleation regime, the found {\it correspondence rule} is fulfilled at all the main stages of crystallization kinetics (crystal nucleation, growth and coalescence). On the example of crystallizing supercooled Lennard-Jones liquid, the rate $g^{-}$ was calculated for the first time at different supercooling levels and for the wide range of nucleus sizes $N\in[n_c;\,40\,n_c]$. It was found that for the whole range of nucleus sizes, the detachment rate $g^{-}$ is only $\approx2$\% less than the attachment rate $g^{+}$. This is direct evidence that the role of the processes that counteract crystallization remains significant at all the stages of crystallization. Based on the obtained results, a kinetic equation was formulated for the time-dependent distribution function of the nucleus sizes, that is an alternative to the well-known kinetic Becker-D\"{o}ring-Zeldovich-Frenkel equation.
\end{abstract}
	

\end{frontmatter}
	
\section{Introduction}

\begin{figure*}[ht!]
	\centering
	\includegraphics[width=1.0\linewidth]{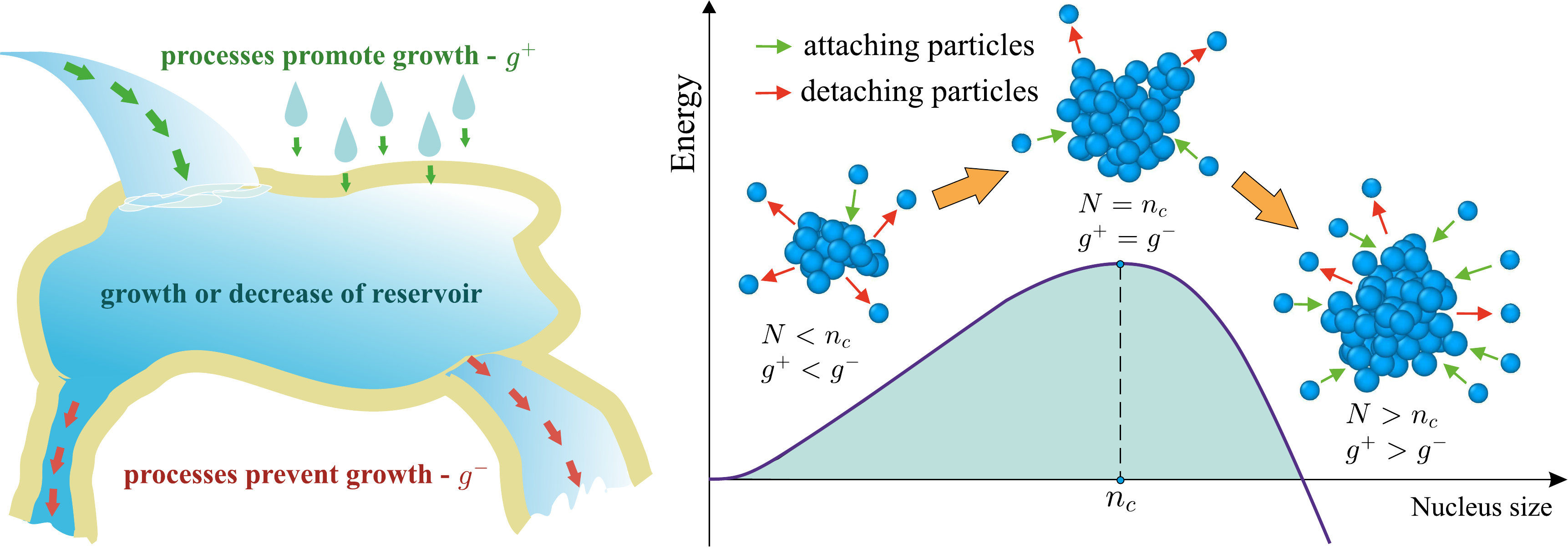}
	\caption{Schematic representation of the processes that promote and prevent the nucleus growth on the example of filling a water reservoir. The process of particle attachment to a nucleus can be compared to the process of filling a reservoir, whereas the detachment of particle from a nucleus is similar to the leakage of water from a reservoir.}
	\label{fig_1}
\end{figure*}

Crystallization of supercooled liquids directly depends on kinetic and thermodynamic factors~\cite{Schmelzer_2010,Zaccone_2013,Brazhkin_2020,Salzmann_2021}. Thermodynamic factors include the interfacial free energy as well as the difference in chemical potentials between two initial (mother) and new (daughter) phases~\cite{Fokin_2016,Skripov_1984,Zeldovich_1942}. Kinetic factors are primarily rate parameters characterizing the frequency of particle ``transitions'' from one phase to another (see Figure~\ref{fig_1}). The number of particles transferred from mother phase to daughter phase per unit time is usually called the attachment rate $g^{+}$~\cite{Zhang_2010,Vetter_2013}. The inverse process associated with the return of particles from new phase to mother phase is characterized by the value $g^{-}$, which is usually denoted as the detachment rate~\cite{Andrianov_2020}. Both quantities $g^{+}$ and $g^{-}$ depend on the size of a new phase nucleus and on its surface area. The larger the surface area, the more probable that a particle located near the nucleus will be able to proceed into the surface layer~\cite{Abyzov_2020}.

In the classical nucleation theory, it is assumed that the shape of the new  phase nuclei is spherical and the nuclei do not contact with each other~\cite{Li_2018,Kashchiev_2000,Kelton_2010,Zaccone_2020}. In this case, the surface layer of the nucleus contains $\sim N^{2/3}$ particles, where $N$ is the number of all particles forming the nucleus. Then the concentration of nuclei with the size $N$ depends only on the frequency of addition/removal particles and it is determined by the master equation of the following form:
\begin{eqnarray}\label{eq_master_equation}
	\frac{dn(N,t)}{dt}&=&g^{+}(N-1)\cdot n_{N-1}(t)+g^{-}(N+1)\cdot n_{N+1}(t) - \nonumber \\ & & - [g^{+}(N)+g^{-}(N)]n_N(t).
\end{eqnarray}
Here, $n_N$ is the concentration of the nuclei with size $N$. Within the framework of the classical nucleation theory, it is assumed that for nuclei of near-critical size (i.e. whose size is $N\sim n_{c}$) the detailed balance condition is satisfied:
\begin{eqnarray}\label{eq_detail_balance}
	g^{+}(N-1)\exp\left(-\frac{\Delta G(N-1)}{k_{B}T}\right)=g^{-}(N)\exp\left(-\frac{\Delta G(N)}{k_{B}T}\right),
\end{eqnarray}
where $\Delta G(N)$ is the minimal free energy required to form a nucleus with size $N$. Condition (\ref{eq_detail_balance}) is usually applied to express $g^{-}(N)$ in terms of the rate $g^{+}(N)$ at $N\approx n_c$, which makes it possible to simplify the solution of kinetic equations for calculation of the crystal growth rate in the framework of such well-known theories as the Wilson-Frenkel theory~\cite{PCCP_2017}, Turnbull-Fisher\cite{Turnbull_1949}, Zeldovich~\cite{Zeldovich_1942}. Thus, the process that counteracts crystallization is indirectly taken into account through the applying the condition (\ref{eq_detail_balance}), and the fact that this condition works only at the beginning of the nucleation regime is usually ignored. Therefore, to obtain the correct solution of the equation (\ref{eq_master_equation}), it is necessary to find a correspondence rule between the kinetic factors $g^{+}(N)$ and $g^{-}(N)$, which will be fulfilled at all the main stages of crystallization kinetics.

Estimation of the kinetic factors $g^{+}(N)$ and $g^{-}(N)$ can be performed with known growth trajectories of crystalline nuclei, for example, obtained using bright-field microscopy~\cite{Wang_2015}, electron microscopy~\cite{Wei_2021}, NMR spectroscopy~\cite{Mashiach_Weissman_2021} or molecular dynamics simulations~\cite{Mendelev_2018,JCP_2020}. As a rule, experimental methods make it possible to obtain only approximate values of the rates $g^{+}(N)$ and $g^{-}(N)$ for crystallization under specific conditions. For example, in the case of crystallization of some molecular glasses, such as ortho-terphenyl, griseofulvin, indomethacin and nifedipine, the structural ordering usually occurs on the surface of liquid solution. The surface crystallization of these molecular glasses is due to the fact that the self-diffusion of surface particles is much higher than the bulk particles. Therefore, the particle attachment and detachment events are easy to determine at the presence of optical contrast between mother and daughter phases~\cite{Mokshin_JETPLett_2019,Wang_2015,Mashiach_Weissman_2021,Zhang_Yu_2016,Huang_Ruan_2017}. Recently, on the basis of NMR spectroscopy data, the dynamics of particles was estimated for the case of the growth of CaF$_{2}$ and SrF$_{2}$ fluoride crystals in water~\cite{Mashiach_Weissman_2021}. In Ref.~\cite{Mashiach_Weissman_2021}, the particle attachment events was fixed by changing the position and intensity of the peaks in NMR spectra, which are responsible for the formation of the core part and the surface of crystallites. Despite the progress achieved in the development of experimental methods for observing the particle attachment and detachment rates, the method of molecular dynamics simulations remains the only way to accurately calculate the quantities $g^{+}(N)$ and $g^{-}(N)$. In molecular dynamics simulations, the trajectories of particles that form both crystalline nuclei and the mother disordered phase are known. This makes it possible to determine with high accuracy the particle attachment/detachment process.

In the present work, we define the kinetic factor $g^{-}$ as a function of the crystalline nucleus size $N$ on the example of the crystallizing supercooled Lennard-Jones liquid. Section $2$ discusses methods for estimating the kinetic factor $g^{-}(N)$. In Section $3$, the obtained results are discussed. The possibility of an accurate estimation of the kinetic factor $g^{-}$ is demonstrated for crystalline nuclei, the size $N$ of which varies over a wide range $N\in[n_c;\,40n_c]$. Section $4$ presents the conclusions. The computational details are given in Section ``Appendix''.

\section{Methods for calculating the kinetic factor $g^{-}$}

\textit{Method of direct calculation.} -- Estimation of the rate $g^{-}$ for nuclei of arbitrary size and shape can be carried out directly from molecular dynamics simulations data. For this purpose, cluster analysis of the simulation results is performed using the local orientational order parameters (see Section ``Appendix'' and Refs.~\cite{JCP_2020,Mickel_Kapfer_2013}), and identification numbers are assigned to each particle. For example, the particles that form crystalline nuclei are assigned the label ``1'', while the particles of the mother phase are denoted by the label ``0''. In this case, the transitions of particles from one phase to another are exactly fixed at each time $t$. Knowing the number of particles $k^{-}$ detached from the nucleus with size $N$ in a small time interval $\Delta t$, we can determine the value $g^{-}(N)$ from the following expression:
\begin{equation}\label{eq_gminus_md}
	g^{-}(N)=\lim_{\Delta t\rightarrow 0}\left\langle\frac{k_{N}^{-}(t)-k_{N}^{-}(t-\Delta t)}{\Delta t}\right\rangle.
\end{equation}  
Here, the time interval $\Delta t$ coincides with the simulation time step. The angle brackets $\langle...\rangle$ mean averaging over different molecular dynamics iterations. Expression (\ref{eq_gminus_md}) does not contain fitting parameters and approximations that ensures high accuracy of the calculated $N$-dependence of the rate $g^{-}$ for a wide range of the size $N$. The applicability of Eq.~(\ref{eq_gminus_md}) also does not depend on the thermodynamic conditions, in which the crystallizing system is located.

\textit{Calculation using known nucleus growth rate and attachment rate.} -- According to the basic definition, the growth rate of new phase nucleus is the difference between the attachment rate $g^{+}(N)$ and the detachment rate $g^{-}(N)$~\cite{Weinberg_2002}: 
\begin{equation}\label{eq_growthrate}
	\vartheta(N)=g^{+}(N)-g^{-}(N).
\end{equation}
This expression corresponds to the case of isotropic growth that occurs with the same manner in all directions. On the other hand, the growth rate $\vartheta(N)$ can be estimated as a function of the size $N$ from the system of equations for the time-dependent growth rate $\vartheta(t)=d\bar{N}(t)/dt$ and the average growth trajectory $\bar{N}(t)$:
\begin{equation}\label{eq_vNdelta}
	\vartheta(\bar{N})=
			\left\{
			\begin{array}{ccc}
		\vartheta = \vartheta(t) \\
		\bar{N} = \bar{N}(t) \\
	\end{array}.
	\right.
\end{equation}
In fact, Eq.~(\ref{eq_vNdelta}) set the equation for $\vartheta(\bar{N})$ in the parametric form. Here, we take $N\equiv\bar{N}$ and assume that the nucleus growth rate $\vartheta$ at the time $t$ must correspond to the $\bar{N}$-sized nucleus at the same time. Then from Eqs.~(\ref{eq_growthrate}) and (\ref{eq_vNdelta}) we find expression for $g^{-}$:
\begin{equation}\label{eq_gminus_rate}
	g^{-}(N)=g^{+}(N)-\vartheta(N).
\end{equation} 
Here, $g^{+}(N)$, $\vartheta(t)$, $\bar{N}(t)$ are the input parameters that can be determined based on experimental data and simulation results~\cite{Mokshin_JETPLett_2019,JCP_2020}. The accuracy of the estimating $g^{-}$ by Eq.~(\ref{eq_gminus_rate}) depends mainly on the correctness of the averaged growth trajectory $\bar{N}(t)$.    

\textit{Theoretical method.} -- Assume that the crystallization of an atomistic system at some stage is a mixture of crystalline nuclei and particles of mother disordered phase. A change in the size of crystalline nuclei occurs only as a result of the attachment and detachment of particles~\cite{Kashchiev_2000}. Then, in accordance with the Turnbull-Fisher kinetic model~\cite{Li_2018,Kelton_2010,Levashov_Ryltsev_2022}, the value of the quantity $g^{-}$ can be approximately determined as follows:
\begin{eqnarray}\label{eq_gminus_theory}
	g^{-}(N)\simeq\gamma K_{N}\exp\left[\frac{\Delta G(N)}{2k_{B}T}\right].
\end{eqnarray} 
Here, the parameter $K_{N}\simeq 4N^{2/3}$ is the number of possible places for detachment of particles on the surface of a spherical nucleus. The quantity $\gamma$ is the frequency parameter, which is a fitting in the present study. For the case of nucleation from the vapor this parameter can be defined via the expression $\gamma\simeq6D/\lambda^{2}$, which is related with the self-diffusion coefficient $D$ and the mean free path of particles $\lambda$~\cite{Kashchiev_2000,Kelton_1991}. According to Eq.~(\ref{eq_gminus_theory}), it is required to expend energy equal to $\Delta G(N)/2$ to detach particles from the nucleus surface.

\section{Estimated kinetic factors for crystallizing supercooled Lennard-Jones liquid}

It is well known that the single-component Lennard-Jones (LJ) system is not capable of forming a stable amorphous phase. The supercooled LJ liquid crystallizes by the mechanism of homogeneous crystal nucleation immediately after the cooling procedure~\cite{Koperwas_2016,Ryzhov_Tareyeva_2020}. Therefore, the supercooled LJ liquid is a convenient system for registering particle attachment/detachment events in relatively short timescales. In the present study, we consider the LJ liquid at the temperatures $T=0.3$, $0.4$ and $0.5\,\epsilon/k_{B}$ corresponding to supercooling levels $66$\%, $55$\% and $43$\%. The growth of the largest crystalline nucleus is taken into account, which makes it possible to perform a more accurate averaging of a single attachment/detachment rate. In this case, the influence of other nuclei on the obtained results will be minimal. The statistical treatment of the results is carried out based on $100$ independent molecular dynamics iterations.

\begin{figure*}[ht!]
	\centering
	\includegraphics[width=1.0\linewidth]{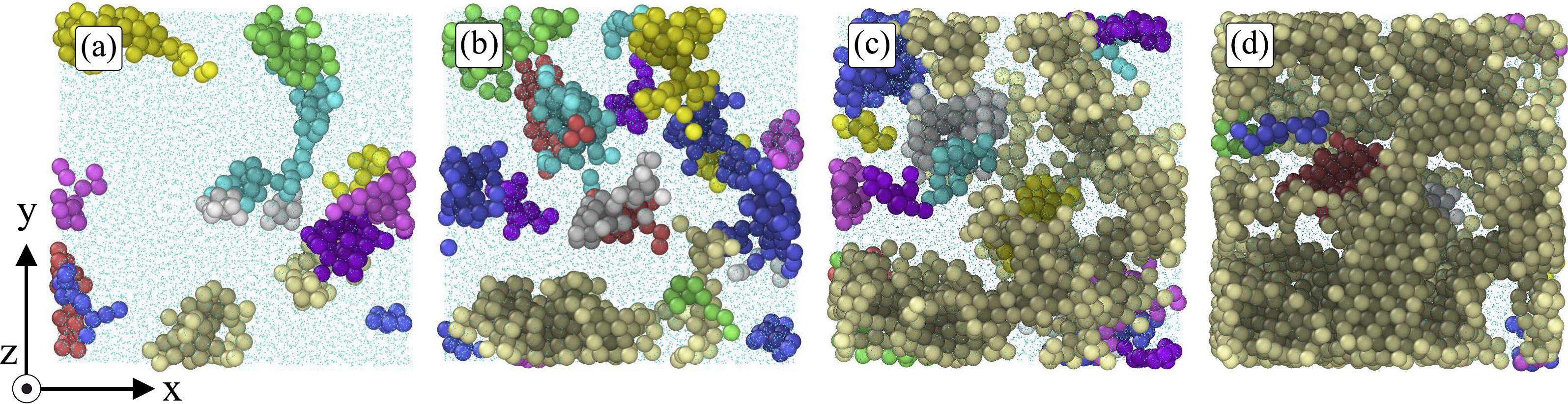}
	\caption{Crystal structure formed in the supercooled LJ liquid at the temperature $T=0.5\,\epsilon/k_{B}$ and at different times $t$: (a) $t=5\,\tau$, (b) $t=25\,\tau$, (c) $t=50\,\tau$ and (d) $t=100\,\tau$. Different colors indicate separated crystalline nuclei.}
	\label{fig_2}
\end{figure*}

Figure~\ref{fig_2} shows that several crystalline nuclei capable of stable growth are formed in the supercooled system. The concentration of such nuclei increases with decreasing temperature due to decrease in the activation energy required to the formation of nuclei of the critical size $n_c$~\cite{Kashchiev_2000,Kelton_2010}. The values of the average critical size $n_c$ and the average nucleation waiting time $\tau_c$ are estimated by the method of inverted averaging the growth trajectories of the largest crystalline nucleus~\cite{PCCP_2017,JCP_2020,Mokshin_JCP_2015,Mokshin_JPCB_2012}. The critical size $n_c$ is about fifty particles at the temperature $T=0.5\,\epsilon/k_{B}$, while at the temperature $T=0.3\,\epsilon/k_{B}$ for the system is more favorable the formation of nuclei of a smaller critical size, where $n_c$ is $40$--$45$ particles. The value of the critical size $n_c$ is determined by the growth curves $\bar{N}(t)$ of crystalline nuclei according to the scheme presented in Ref.~\cite{JCP_2020}; the found values of the critical size $n_c$ are given in Table~\ref{table_1}. At the initial stage of crystallization, the nuclei grow mainly due to attachment of single particles, while at the final stage of crystallization the nuclei grow by coalescing with smaller crystallites according to the restructuring/absorption mechanism~\cite{Galimzyanov_Ladyanov_2019}. We find that under the considered thermodynamic ($p$, $T$)-conditions, the contact of crystalline nuclei is weakly pronounced if their sizes do not exceed $\sim20\,n_c$.

\begin{figure*}[ht!]
	\centering
	\includegraphics[width=1.0\linewidth]{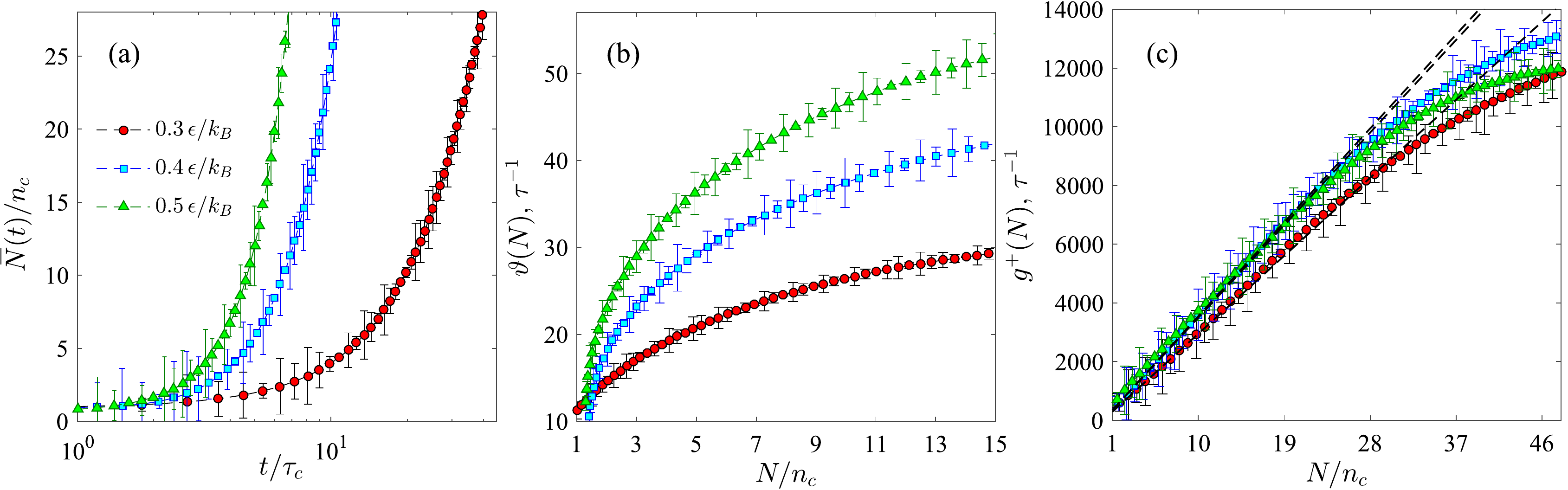}
	\caption{(a) Reduced growth curves $\bar{N}(t)/n_c$ of the largest nucleus versus the reduced time $t/\tau_c$ at various temperatures. The values of the critical size $n_c$ and the nucleation waiting time $\tau_c$ are calculated based on simulation data and presented in Table~\ref{table_1}. (b) Growth rate $\vartheta$ calculated by Eq.~(\ref{eq_vNdelta}) based on molecular dynamics simulation data. (c) $N/n_c$-dependence of the attachment rate $g^{+}$ obtained by direct calculation~\cite{JCP_2020}. The dashed curves show the results of Eq.~(\ref{eq_gplus_fit}).}
	\label{fig_3}
\end{figure*}

Figure~\ref{fig_3}(a) shows that the stable growth of the quantity $\bar{N}(t)$ begins only after the appearance of the critically-sized nucleus $n_c$ during the time $\tau_c$ [see Table~\ref{table_1}]. The displacement of the $\bar{N}(t)$-curves towards longer times $t$ with decreasing temperature is associated with the suppression in crystallization due to an increase in viscosity. As shown in Figure~\ref{fig_3}(b), the $N$-dependence of the nucleus growth rate $\vartheta$ also decreases with decreasing temperature. The value of the growth rate $\vartheta$ demonstrates a pronounced dependence on the nucleus size in the range $N\in[n_c;\,15\,n_c]$: the larger the nucleus, the higher its growth rate. The growth of the nucleus slows down after it reaches the size $N\approx 20n_{c}$, which is indicated by the saturation of the $\vartheta(N)$-curves. This is mainly due to the fact that the attachment rate and the detachment rate take similar values when the largest nucleus begins to absorb small crystallites by the formation of a common crystal lattice~\cite{Galimzyanov_Ladyanov_2019}. Figure~\ref{fig_3}(c) shows that the attachment rate $g^{+}$ rapidly increases with the nucleus size and reaches saturation only at the final stage of crystallization. The dependence $g^{+}(N)$ for the size range $N\in[n_c;\,20\,n_c]$ is reproduced by the power-law
\begin{equation}\label{eq_gplus_fit}
	g^{+}(N)=g_{n_c}^{+}\left(\frac{N}{n_c}\right)^{\xi^{+}}.
\end{equation}   
Here, $g_{n_c}^{+}$ is the attachment rate of particles to the nucleus of the critical size $n_c$ [see Table~\ref{table_1}]. The exponent $\xi^{+}$ characterizes the growth regime and takes the value $\xi^{+}\simeq(1.0\pm0.005)$. This value corresponds to the limiting case, when the attachment rate increases linearly with the nucleus size~\cite{PCCP_2017}. This growth regime is specific for nuclei with a complex surface geometry, the large area of which contributes to a rapid increase in the number of attached particles. In the case of growth of the nucleus whose shape is close to spherical (that is possible, for example, at low supercooling levels), the linear regime becomes less pronounced and the $N$-dependence of the quantity $g^{+}$ mainly follows the power law $g^{+}(N)\sim N^{2/3}$.  

\begin{table}[ht!]
	\centering
	\caption{Characteristics of the crystallizing system: temperature $T$; critical size $n_c$; average time $\tau_c$ for formation of critically-sized nucleus; attachment rate $g_{n_c}^{+}$ for critically-sized nucleus; detachment rate $g_{n_c}^{-}$ for the nucleus of critical size $n_c$; values of the parameters $\gamma$ and $\Delta G/(k_{B}T)$ in Eq.~(\ref{eq_gminus_theory}).}\label{table_1}
	\begin{tabular}{ccccccc}
		\hline\hline
		$T$, $\epsilon/k_B$ & $n_c$ & $\tau_c$, $\tau$ & $g_{n_c}^{+}$, $\tau^{-1}$ & $g_{n_c}^{-}$, $\tau^{-1}$ & $\gamma$, $\tau^{-1}$ & $\Delta G/(k_{B}T)$ \\
		\hline
		$0.5$  & $52\pm2$ & $11\pm1.3$ 		& $355\pm40$ & $365\pm40$ & $6.3$ & $2.0$ \\ 
		$0.4$  & $48\pm2$ & $4.5\pm0.6$ 	& $352\pm35$ & $350\pm35$ & $7.8$ & $1.8$ \\
		$0.3$   & $45\pm3$ & $1.5\pm0.2$	& $295\pm25$ & $295\pm25$ & $8.2$ & $1.5$   \\
		\hline\hline
	\end{tabular}
\end{table}

\begin{figure*}[ht!]
	\centering
	\includegraphics[width=1.0\linewidth]{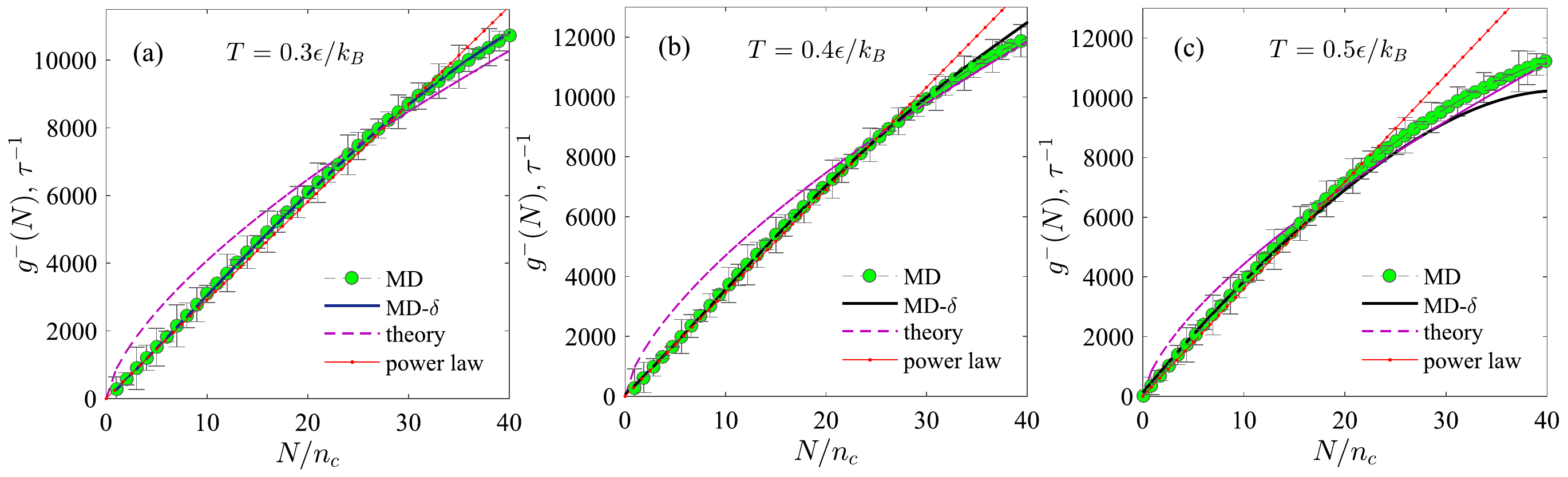}
	\caption{Dependence of the kinetic factor $g^{-}$ on the nucleus size $N$: (a) at the temperature $T=0.3\,\epsilon/k_{B}$, (b) $T=0.4\,\epsilon/k_ {B}$ and (c) $T=0.5\,\epsilon/k_{B}$. The results of Eqs.~(\ref{eq_gminus_md}) and (\ref{eq_gminus_rate}) are denoted as MD and MD-$\delta$ respectively. The dashed lines show the result of Eq.~(\ref{eq_gminus_theory}) at the values of the fitting parameters $\gamma$ and $\Delta G/(k_{B}T)$ given in Table~\ref{table_1}. The dotted curves are the results of Eq.~(\ref{eq_gminus_fit}).}
	\label{fig_4}
\end{figure*}

Figure~\ref{fig_4} shows the $N/n_c$-dependences of the detachment rate $g^{-}$ at different temperatures calculated directly by expression (\ref{eq_gminus_md}) and compared with the results of Eqs.~(\ref{eq_gminus_rate}) and (\ref{eq_gminus_theory}). At the considered temperatures, these dependencies are increasing functions and have the same form as the $N/n_c$-dependences of the attachment rate $g^{+}$. A good agreement between the results of direct calculation and the data obtained by Eq.~(\ref{eq_gminus_rate}) is due to the fact that the kinetic factors $g^{+}$ and $g^{-}$ take close values. In this case, the attachment rate $g^{+}$ in its values only slightly exceeds the values of the detachment rate $g^{-}$.

As can be seen from Figure~\ref{fig_4}, good agreement between the simulation data and theoretical results is observed only at the temperature $T=0.5\,\epsilon/k_{B}$ corresponding to moderate supercooling. At other considered temperatures, the theory produces overestimated values of the quantity $g^{-}$ for nuclei with sizes less than $20\,n_c$. This is due to the fact that in expression (\ref{eq_gminus_theory}) the value $g^{-}$ is related to the nucleus  size through the law $g^{-}(N)\propto N^{2/3}$, which implies the  detachment of particles from the nucleus surface equally probable in all directions. The small size of nuclei and their high concentration at low temperatures, for example, at the temperatures $T=0.3\,\epsilon/k_{B}$ and $T=0.4\,\epsilon/k_{B}$, contribute to unevenly growth of crystallites. Therefore, more accurate theoretical calculations with Eq.~(\ref{eq_gminus_theory}) can be carried out only for the case of the so-called kinetically limited growth, which occurs at low levels of supercooling as well as at the stage of coalescence of crystallites~\cite{Galimzyanov_Ladyanov_2019,Slezov_2009}.

The calculated $N/n_c$-dependencies of the detachment rate $g^{-}$ are reproduced by the power law
\begin{equation}\label{eq_gminus_fit}
	g^{-}(N)=g_{n_c}^{-}\left(\frac{N}{n_c}\right)^{\xi^{-}},
\end{equation}  
which is similar to Eq.~(\ref{eq_gplus_fit}) for the attachment rate $g^{+}$. In Eq.~(\ref{eq_gminus_fit}), the exponent takes the value $\xi^{-}\simeq(0.995\pm0.005)$ and does not depend on the thermodynamic state of the system. It is noteworthy that the found value of the parameter $\xi^{-}$ is close to the value of the parameter $\xi^{+}$: the difference between these parameters is $\xi^{+}-\xi^{-}\simeq0.005$. The value of the parameter $\xi^{-}$ is close to unity, which indicates the linear $N/n_c$-dependence of the quantity $g^{-}$. As can be seen from Figure~\ref{fig_4}, at the temperature $T=0.3\,\epsilon/k_{B}$, the linear region corresponds to the range $N\in[n_c;\,25n_c]$, while at $T= 0.5\,\epsilon/k_{B}$ such the region covers the narrow range $N\in[n_c;\,15n_c]$. The narrowing of the linear region with increasing temperature can be associated with the transition to the scenario of kinetically limited growth, when the detachment rate $g^{-}$ depends mainly on the shape and surface area of the nucleus.

The presence of a small positive difference between $\xi^{+}$ and $\xi^{-}$ is due to the fact that the attachment rate $g^{+}$ only slightly exceeds the detachment rate $g^{-}$. We assume that the parameters $\xi^{+}$ and $\xi^{-}$ will always take close values for crystallizing supercooled liquids, where the formation and growth of nuclei occurs due to the competition between the particle attachment and detachment processes. Then from Eqs.~(\ref{eq_gplus_fit}) and (\ref{eq_gminus_fit}) we get the following correspondence rule:
\begin{equation}\label{eq_gplus_gminus_fit}
	\frac{g^{+}(N)}{g^{-}(N)}\simeq\left(\frac{N}{n_c}\right)^{\xi^{+}-\xi^{-}},
\end{equation}  
which relates the quantities $g^{+}(N)$ and $g^{-}(N)$. This rule allows us to express $g^{+}(N)$ in terms of the quantity $g^{-}(N)$ for nuclei of arbitrary size. It follows from Eq.~(\ref{eq_gplus_gminus_fit}) that the detachment rate $g^{-}$ is only $\approx2$\% less than the attachment rate $g^{+}$ for the whole range of nucleus sizes. Thus, this is direct evidence that the role of processes that counteract crystallization remains significant at all stages of crystallization.

Taking into account that $g_{n_c}^{+}=g_{n_c}^{-}$ for the critically-sized nucleus, from Eqs.~(\ref{eq_master_equation}), (\ref{eq_gplus_fit}) and (\ref{eq_gminus_fit}) we find the following master equation for calculating the rate of change in the concentration of nuclei with size $N$:
\begin{eqnarray}\label{eq_master_equation_1}
	\frac{dn(N,t)}{dt} &=& g_{n_c}^{+}\left\{\left(\frac{N-1}{n_c}\right)^{\xi^{+}}n_{N-1}(t)+\left(\frac{N+1}{n_c}\right)^{\xi^{-}}n_{N+1}(t)\right\} 
	- \nonumber \\ & & - g_{n_c}^{+}\left[\left(\frac{N}{n_c}\right)^{\xi^{+}}+\left(\frac{N}{n_c}\right)^{\xi^{-}}\right]n_{N}(t).
\end{eqnarray}
In a specific case, for the crystallization regime at $N>>n_{c}$ we can assume that $(N+1)/n_c\simeq N/n_{c}$ and $(N-1)/n_c\simeq N/n_{c}$.  Then Eq.~(\ref{eq_master_equation_1}) can be rewritten in the following form:
\begin{eqnarray}\label{eq_master_equation_2}
\frac{dn(N,t)}{dt} &=& g_{n_c}^{+}\left\{\left(\frac{N}{n_c}\right)^{\xi^{+}}[n_{N-1}(t)-n_{N}(t)]\right\} + \nonumber \\ & & +g_{n_c}^{+}\left\{\left(\frac{N}{n_c}\right)^{\xi^{-}}[n_{N+1}(t)-n_{N}(t)]\right\}
\end{eqnarray}
or
\begin{eqnarray}\label{eq_master_equation_3}
\frac{dn(N,t)}{dt} = g_{n_c}^{+}\left\{\left(\frac{N}{n_c}\right)^{\xi^{-}}\frac{\partial f_{N+1}}{\partial N}-\left(\frac{N}{n_c}\right)^{\xi^{+}}\frac{\partial f_{N-1}}{\partial N}\right\},
\end{eqnarray}
where
\begin{eqnarray}\label{eq_master_equation_4}
\frac{\partial f_{N+1}}{\partial N}=\frac{n_{N+1}(t)-n_{N}(t)}{(N+1)-N},
\end{eqnarray}
\begin{eqnarray}\label{eq_master_equation_5}
\frac{\partial f_{N-1}}{\partial N}=\frac{n_{N}(t)-n_{N-1}(t)}{N-(N-1)}.
\end{eqnarray}
Thus, equation (\ref{eq_master_equation_1}) is an alternative to the well-known kinetic Becker-D\"{o}ring-Zeldovich-Frenkel equation~\cite{Zeldovich_1942,Becker_Doring_1935,Frenkel_1939}.

\section{Conclusions}
In the present work, for the first time, the process inherent in crystallization, which determines it and counteracts the transition of particles from the liquid phase to the crystalline one, is considered in detail. Various methods for estimation of the detachment rate $g^{-}$ as a function of the crystalline nucleus size $N$ were presented: the method of direct calculation based on the results of molecular dynamics simulations [expression (\ref{eq_gminus_md})]; the method based on the known nucleus growth rate $\vartheta(N)$ and the attachment rate $g^{+}(N)$ [expression (\ref{eq_gminus_rate})]. On the example of homogeneous crystal nucleation in the supercooled LJ liquid, it is shown that these methods allow one to correctly calculate the dependence of the kinetic factor $g^{-}$ on the nucleus size for the wide range of sizes $N\in[n_c;\,40n_c]$: from the stage of nucleation of stable crystallites to the final stage of system crystallization. Such results have not been previously reported in the scientific literature.

The effects associated with a certain finite size of the simulation cell -- the so-called ``finite size effects'' -- can indeed affect the results for the kinetic factors $g^{+}$ and $g^{-}$. The role of these effects depends on supercooling level of the system. In addition, these effects become more pronounced at the stages of crystallization, at which the linear dimension of growing nuclei become comparable to the size of the simulation cell, and when the concentration of nuclei becomes significant. More specifically, the finite size effects directly determine the start of saturation for the size-dependent detachment rate. However, in this study, the transition of the rate factors to the saturation regime is not considered, but the main attention is paid to the general regime associated with a monotonic increase in rates $g^{+}$ and $g^{-}$ with the nucleus size $N$. Thus, the finite size effects are naturally ignored in the study.

From a physical point of view, the values of the rate coefficients $g^{+}$ and $g^{-}$ are determined by common kinetic and thermodynamic motives, which manifest themselves in the values of viscosity, interfacial free energy, etc. In addition, the values of these coefficients can also be determined by structural and dynamic heterogeneities that arise in a liquid (even simple) at strong supercoolings as well as by the effects associated with the interactions of growing nuclei of a new (crystalline) phase. All these effects can have a significant impact on crystallization kinetics and they define the values of $g^{+}$ and $g^{-}$. In the given study, the rate coefficients $g^{+}$ and $g^{-}$ were found directly on the basis of the corresponding registered events related with attachment and detachment of monomers. Thus, the $g^{+}$ and $g^{-}$ values obtained using the applied calculation scheme contain information about all the effects mentioned above.

The results of theoretical calculations are revealed that the $N$-dependence of the rate $g^{-}$ is not reproduced by any general equation, for example, as Eq.~(\ref{eq_gminus_theory}), due to the mixing of different crystal growth scenarios. The scenario of kinetically limited growth with $g^{-}(N)\sim N^{2/3}$ is realized at the stage of formation of critically-sized nuclei. The realization of this scenario was also discussed earlier in the works of M.I.~Mendelev and co-authors on the example of calculating the kinetic factor $g^{+}$ in the case of crystallization of amorphous pure Ni and Ni-Al alloys~\cite{Mendelev_2018}. The following scenario occurs when the detachment rate increases linearly with the increasing nucleus size at the stage of stable growth of nuclei. This scenario is quite expected in the case of deep supercooling due to unevenly growth of crystalline nuclei~\cite{Slezov_2009}. At the stage of coalescence, transition to the scenario close to kinetically limited growth is observed due to the rearrangement of the crystal structure and the consolidation of nuclei~\cite{Galimzyanov_Ladyanov_2019}. In this case, the detachment and attachment kinetics of particles becomes equally probable in all directions. It is important to note that the kinetic factors $g^{+}$ and $g^{-}$ show the same $N$-dependence regardless of the system supercooling and these factors are related by the found correspondence rule. This correspondence rule can be used to obtain more accurate solution of kinetic equations as applied to the case of crystallization of supercooled or glassy systems with more complex interatomic interactions, for example, such as metallic alloys, molecular systems and colloidal solutions. 

\section*{Acknowledgment}
\noindent The work was supported by the Russian Science Foundation (project No. 19-12-00022). DTY and AVM are grateful to the Foundation for the Development of Theoretical Physics and Mathematics ``Basis'', which supported part of the work related to the development of computational algorithms.

\section*{Appendix: computation details}

We consider the supercooled Lennard-Jones (LJ) liquid, which spontaneously crystallizes according to the homogeneous scenario. The simulation is performed using the Lennard-Jones potential~\cite{Ryzhov_Tareyeva_2020,Stillinger_2001}:
\begin{equation}\label{eq_LJ}
U(r_{ij})=4\epsilon\left[\left(\frac{\sigma}{r_{ij}}\right)^{12}-\left(\frac{\sigma}{r_{ij}}\right)^{6}\right].
\end{equation}
Here, $\sigma$ is the effective diameter of the particle (atom or molecule), $\epsilon$ is the depth of the potential well, $r_{ij}$ is the distance between particles $i$ and $j$. The cutoff radius is $r_{cut}=2.5\,\sigma$ and we have  $U(r_{ij})=0$ at $r_{ij}\geq r_{cut}$. For the considered LJ system we have $\sigma=1$ and $\epsilon=1$. Particles in the count $13\,500$ are located inside the cubic simulation cell with side lengths $L_x\approx L_y\approx L_z\approx 15\,\sigma$. Periodic boundary conditions apply in all directions. Integration of the equations of motion is carried out with the time step $\Delta t=0.01\,\tau$, where $\tau$ is the time unit. 

Liquid samples of the system were brought to a thermodynamic equilibrium at the temperature $T=2.5\,\epsilon/k_B$ on the isobar $p=2.0\,\epsilon/\sigma^3$ ($k_{B}$ is the Boltzmann constant). Supercooled samples were obtained by rapid cooling of the liquid to the temperatures $T=0.3$, $0.4$ and $0.5\,\epsilon/k_B$ at the isobar $p=2.0\,\epsilon/\sigma^3$. The cooling rate is $0.02\,\epsilon/(k_{B}\tau)$, which in the case of a real system corresponds to $\sim1\times10^{12}$\,K/s. At the considered temperatures, the supercooling of the system is $(T_m-T)/T_m = 0.66$, $0.55$, and $0.43$ respectively. Here, the melting temperature of the system is $T_{m}\simeq0.88\,\epsilon/k_{B}$ at the pressure $p=2.0\,\epsilon/\sigma^3$ (see the phase diagram of the LJ system~\cite{Travesset_2014}).

Identification of the crystalline structures in supercooled system is performed by the method of cluster analysis based on computation of orientational order parameters~\cite{Mickel_Kapfer_2013,Wolde_1996,Steinhardt_Nelson_1983,Lechner_2008}. For the single-component LJ system, just a single parameter $q_6$ is sufficient to recognize correctly the particles that belong to crystalline phases typical for the LJ system~\cite{Steinhardt_Nelson_1983}. This parameter is calculated by the expression:
\begin{equation}\label{eq_bop}
q_{6}(i)=\left(\frac{4\pi}{13}\sum_{m=-6}^{6}|q_{6m}(i)|^{2}\right)^{1/2},
\end{equation}
where
\begin{equation}\label{eq_bop_1}
q_{6m}(i)=\frac{1}{n_{b}(i)}\sum_{j=1}^{n_{b}(i)}Y_{6m}(\theta_{ij},\phi_{ij}).
\end{equation}
Here, $n_{b}(i)$ is the number of neighbors for the $i$th particle, $Y_{6m}(\theta_{ij},\phi_{ij})$ are the spherical harmonics with the polar $\theta_{ij}$ and azimuthal $\phi_{ij}$ angles. To recognize crystalline structures, the ten Wolde-Frenkel condition is applied~\cite{Wolde_1996}:
\begin{equation}\label{eq_wf_cond1}
0.5<\left|\sum_{m=-6}^{6}\bar{q}_{6m}(i)\bar{q}_{6m}^{*}(j)\right|\leq 1,
\end{equation}
where
\begin{equation}\label{eq_wf_cond2}
\bar{q}_{6m}=\frac{1}{n_{b}(i)}\frac{q_{6m}(i)}{\sqrt{\sum_{m=-6}^{6}|q_{6m}(i)|^{2}}}.
\end{equation}
The $i$th particle is considered as in a crystalline phase if this particle has four or more crystal-like bonds with their own ``neighbors''.

\end{document}